\documentclass[conference]{IEEEtran}
\IEEEoverridecommandlockouts
\usepackage{cite}
\usepackage{amsmath,amssymb,amsfonts}
\usepackage{algorithmic}
\usepackage{balance}
\usepackage{algorithm}
\usepackage{amsmath}
\usepackage{array}
\usepackage{algorithmic}
\usepackage{graphicx}
\usepackage{textcomp}
\usepackage{listings}
\usepackage{xcolor}
\usepackage{subfig}
\usepackage{scalerel} 
\usepackage{tikz}
\usepackage{pgfplots}
\usetikzlibrary{arrows}
\usetikzlibrary{shapes}

\pgfplotsset{width=2cm,compat=1.8}
\usetikzlibrary{svg.path}
\usetikzlibrary{patterns}
\def\BibTeX{{\rm B\kern-.05em{\sc i\kern-.025em b}\kern-.08em
    T\kern-.1667em\lower.7ex\hbox{E}\kern-.125emX}}
    \definecolor{orcidlogocol}{HTML}{A6CE39}
\tikzset{
  orcidlogo/.pic={
    \fill[orcidlogocol] svg{M256,128c0,70.7-57.3,128-128,128C57.3,256,0,198.7,0,128C0,57.3,57.3,0,128,0C198.7,0,256,57.3,256,128z};
    \fill[white] svg{M86.3,186.2H70.9V79.1h15.4v48.4V186.2z}
                 svg{M108.9,79.1h41.6c39.6,0,57,28.3,57,53.6c0,27.5-21.5,53.6-56.8,53.6h-41.8V79.1z M124.3,172.4h24.5c34.9,0,42.9-26.5,42.9-39.7c0-21.5-13.7-39.7-43.7-39.7h-23.7V172.4z}
                 svg{M88.7,56.8c0,5.5-4.5,10.1-10.1,10.1c-5.6,0-10.1-4.6-10.1-10.1c0-5.6,4.5-10.1,10.1-10.1C84.2,46.7,88.7,51.3,88.7,56.8z};
  }
}

\newcommand\orcidicon[1]{\href{https://orcid.org/#1}{\mbox{\scalerel*{
\begin{tikzpicture}[yscale=-1,transform shape]
\pic{orcidlogo};
\end{tikzpicture}
}{|}}}}

\usepackage{hyperref}

\begin{document}

\title{A Graph-based Approach to
Human Activity Recognition}

 \author{Thomas Peroutka\IEEEauthorrefmark{1}, Ilir Murturi\IEEEauthorrefmark{1} \orcidicon{0000-0003-1990-0431}, Praveen Kumar Donta\IEEEauthorrefmark{2}\orcidicon{0000-0002-8233-6071}, and Schahram Dustdar\IEEEauthorrefmark{1} \orcidicon{0000-0001-6872-8821} \\ 
 \IEEEauthorblockA{\IEEEauthorrefmark{1}{Distributed Systems Group}, {TU Wien}, Vienna 1040, Austria. }
  \IEEEauthorrefmark{2}{Department of Computer and Systems Sciences}, {Stockholm University}, Stockholm 16425, Sweden.
 }
 \IEEEoverridecommandlockouts
\IEEEpubid{\makebox[\columnwidth]{\hfill} \hspace{\columnsep}\makebox[\columnwidth]{ }}
\maketitle
\IEEEpubidadjcol
\begin{abstract}
Advanced wearable sensor devices have enabled the recording of vast amounts of movement data from individuals regarding their physical activities. This data offers valuable insights that enhance our understanding of how physical activities contribute to improved physical health and overall quality of life. Consequently, there is a growing need for efficient methods to extract significant insights from these rapidly expanding real-time datasets. This paper presents a methodology to efficiently extract substantial insights from these expanding datasets, focusing on professional sports but applicable to various human activities. By utilizing data from Inertial Measurement Units (IMU) and Global Navigation Satellite Systems (GNSS) receivers, athletic performance can be analyzed using directed graphs to encode knowledge of complex movements. Our approach is demonstrated on biathlon data and detects specific points of interest and complex movement sequences, facilitating the comparison and analysis of human physical performance.
\end{abstract}


\section{Introduction} 
\label{intro}
Throughout human evolution, our bodies and brains learned the ever-increasing motion complexity. Even decades into the computer and machine-powered area, it is still hard for machines to solve supposedly easy tasks like manipulating a Rubik's cube \cite{OpenAI-1910-07113}. This means humans and all living things have the unique ability to move their bodies accurately and efficiently. With more and more wearable sensors available, such movements can be measured precisely and gain insights into efficiency. The wearable technology market is predicted to grow at a compound annual growth rate of 14.6\% from 2023 to 2030 \cite{GVR2023}. This includes using standard hardware like "Cardiovascular Monitoring Using Earphones and a Mobile Device" \cite{Poh2011} or "A Wearable Sensor for Measuring Sweat Rate" \cite{Sal2010}. There are even sensors that act as little radar to detect human activities based on millimeter-waves \cite{Yu2022}. All these sensors generate a large amount of data with a common objective: detecting physical activities and extracting insights. This field of study is known as Human Activity/Action Recognition (HAR) \cite{lara2012survey}.

The HAR field researches how to identify and understand human activities using technology. Despite the progress made in this field, crucial aspects should be addressed to significantly transform how individuals interact with mobile devices and other devices. In the last few years, there have been extensive studies in multiple areas, such as 3D Convolutional Neural Networks (CNN) for HAR.  Ji et al. \cite{Ji2010} demonstrated how image-based activity recognition is possible. In contrast, Bia et al. \cite{Bia2019} showcases how a wearable sensor and a CNN can detect human activity like walking, sitting, or climbing stairs. Another approach is a statistical analysis of given data and trying to classify the activities \cite{Deh2018}. The current research aims to interpret datasets without prior knowledge, meaning that they do not describe what movements look like to the system. Meanwhile, most of these approaches are resource-intensive (using neural networks) or are difficult to set up. Therefore, our goal in this paper is to use domain-specific knowledge (i.e., anatomy of a movement sequence) obtained by experts (e.g., trainers) and encoded into a directed graph to quickly and reliably detect motion sequences (i.e., movements). Furthermore, we emphasize using existing algorithms, which make analyzing big datasets efficient and resource-friendly.  Nevertheless, our approach is limited to human activity recognition; however, it can also be used in any time data series.

In this paper, we propose an approach that aims to represent complex athletes' movements as a directed graph. The goal is to find an efficient method for identifying critical points in a multi-sensor dataset and detecting complex movements. This approach is versatile and can be applied to any movement. For illustration, we will focus on biathlon, a demanding sport that combines two very different activities: fast skiing on various terrains and precise target shooting while stationary. Speed is crucial in biathlon, so finding the most effective and efficient body movement techniques is essential. This involves determining the best approach for uphill, flat, or downhill sections and maximizing shooting accuracy. Moreover, we aim to identify a data structure capable of capturing body movements and developing an algorithm that can efficiently detect these movements using multiple wearable sensors. Ultimately, we will obtain performance indicators specific to different movements that can be compared with others. This will enable professional athletes, sports enthusiasts, and rehabilitation patients to enhance their physical movements. Note that this work focuses on the technological challenges associated with detecting user-defined motion sequences, and any sports science insights provided in this article are purely for illustrative purposes and may not be scientifically grounded.


The remaining sections are structured as follows. 
A motivation example and related work are presented in Section \ref{motiv1}. Section \ref{section2} presents the proposed approach and an example to illustrate how to represent a sequence of movements in a directed graph.  Evaluation results are
discussed in Section \ref{chap:experiments}. Lastly, we conclude our discussion with possible future actions in Section \ref{conclusion}.


\section{Motivation and Related Work} 
 \label{motiv1}
\subsection{Motivation Scenario}
Understanding and improving athlete performance is critical in professional sports (i.e., especially in disciplines like biathlon). Biathlon combines cross-country skiing and rifle shooting, demanding peak physical condition and precision. Coaches and sports researchers constantly seek new methods to analyze and optimize the performance of athletes. Imagine a biathlon team preparing for the winter sports. The team's performance analytics division provides detailed insights into each athlete's performance to identify strengths and areas for improvement.  

Analyzing transitions between race stages helps to identify athlete delays and fatigue signs (e.g., observing how athletes manage the shift from starting to tackling the uphill challenge, and then from the uphill to the shooting range). Traditional performance analysis methods are time-consuming and often lack the granularity to make such fine-tuned adjustments. Using IMUs and GNSS receivers, the responsible team can collect detailed movement data from athletes during training and competitions. Directed graphs allow encoding complex movement sequences, which in return allows the team to (i) capture detailed movement patterns, (ii) identify possible performance bottlenecks, (iii) optimize training, or (iv) compare athlete performances.  

\subsection{Related Work}
HAR is a rapidly evolving field with various approaches to tracking and interpreting human actions \cite{li2020survey}. A common methodology involves analyzing images and videos to detect and extract human figures and their movements \cite{Jalal2014, bukht2024review, kumar2024human}. The authors present a real-time system for 3D human pose estimation, tracking, and recognition from RGB-D video sequences using a generative structured framework. Kaya et al. \cite{kaya2024human} proposed a new 1D-CNN-based deep learning approach for sensor-based HAR using raw accelerometer and gyroscope data. The proposed work showed the impact of using individual sensor data versus combined data while finding that the model performed better with concurrent sensor data. Liu et al. \cite{liu2021bayesian} applied the Bayesian structural time series framework to biomedical sensor data, demonstrating its ability to accurately assess the significance of health interventions while accounting for complex covariate structures. The developed tool called MhealthCI processes and registers diverse biomedical data for personalized medicine applications. Kumar et al. \cite{kumar2023deep} propose a novel Deep-HAR model that combines CNNs for feature extraction and Recurrent Neural Networks (RNNs) for pattern recognition in time-series data. On the contrary to these works, the innovative technique utilizes WiFi technology to ascertain human presence without the need for cameras or sensors \cite{Zhang2024}.

Contrary to the above-mentioned works, our approach focuses on the most common way to use inertial sensors in wearable devices.  The majority of research is concerned with classification, but very little is done in extracting more than just classes from datasets. Therefore, we aim to take it a step further and not only classify activities (e.g., smartwatches or smartphones classify human activities \cite{Ramanujam2021}) but also be able to extract user-defined performance metrics like forces applied to the body during certain tasks, etc. Our proposed approach requires domain-specific knowledge, requiring an understanding of how to break down a complex movement into key points. In contrast to other research works, our proposed graph-based method combined with domain-specific knowledge enables complex movement detection in a resource-efficient manner. Furthermore, our approach generates explainable results (meaning it is possible to reason (i.e., results are traceable and how they were calculated), which most approaches, especially ones using neural networks, lack. The advantages of our approach arise from the combination of a strict rule set driven by domain-specific knowledge and the expressive capabilities of graphs. This unique combination provides key benefits to our approach. 

\section{The proposed approach} 
\label{section2}
\subsection{An overview of the approach}
\begin{figure}[h]
    \centering
    \includegraphics[width=\columnwidth]{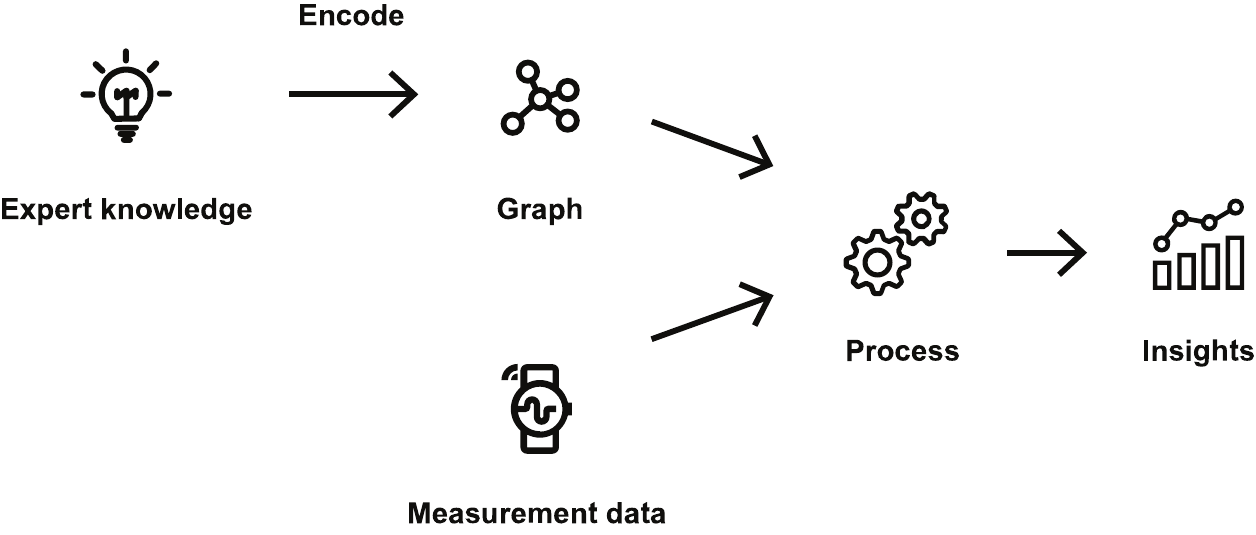}
    \caption{An overview of the proposed approach.}
    \label{fig:approach-overview}
\end{figure}

In Figure \ref{fig:approach-overview}, we outline the concept of the proposed approach. The first step is to break down a specific movement into smaller parts that make up the movement. This is done with the help of an expert (i.e., typically trainers or sports scientists). This step, e.g., includes trainers setting up virtual gates on a map or defining triggers for acceleration data. The temporal dependency between those parts is encoded into a directed graph. Movements are recorded with wearable devices featuring many sensors (e.g., IMU, GNSS, heart rate sensor, etc.). The data is then processed by our proposed approach (i.e., described in the next sections) and trainers receive insights about the movements of interest.

\subsection{Encoding complex movements into a directed graph} \label{encoding}
\begin{figure} 
        \centering
            \centering
            \includegraphics[width=\columnwidth]{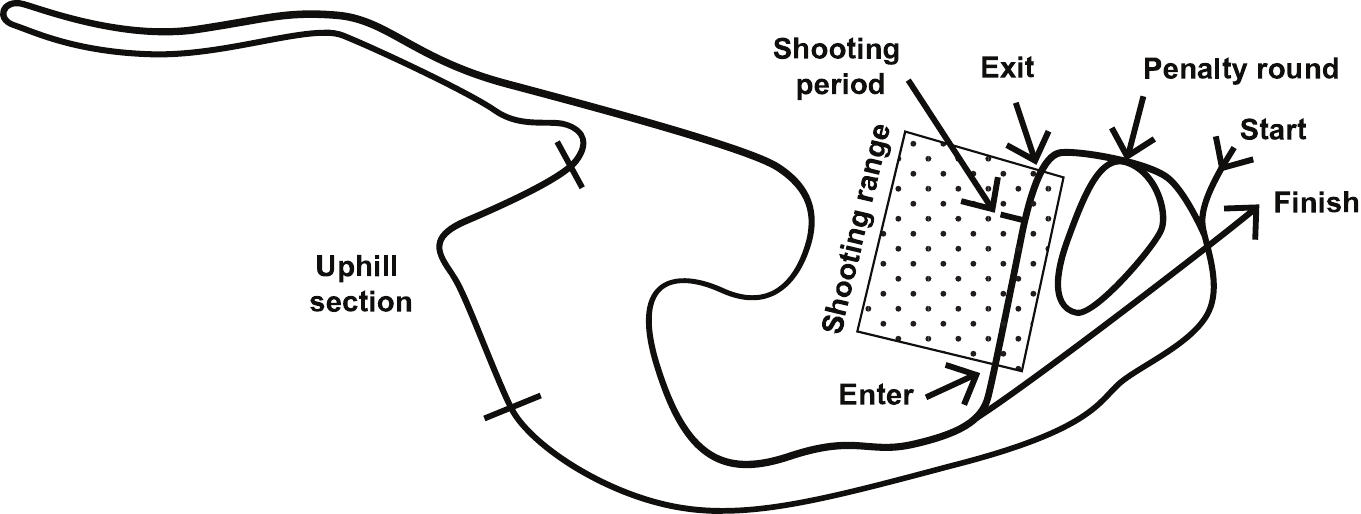}
            \caption{A typical biathlon racetrack layout.}
            \label{fig:racetrack-schema}
        \end{figure}

\begin{figure}
            \centering
            \includegraphics[width=\columnwidth]{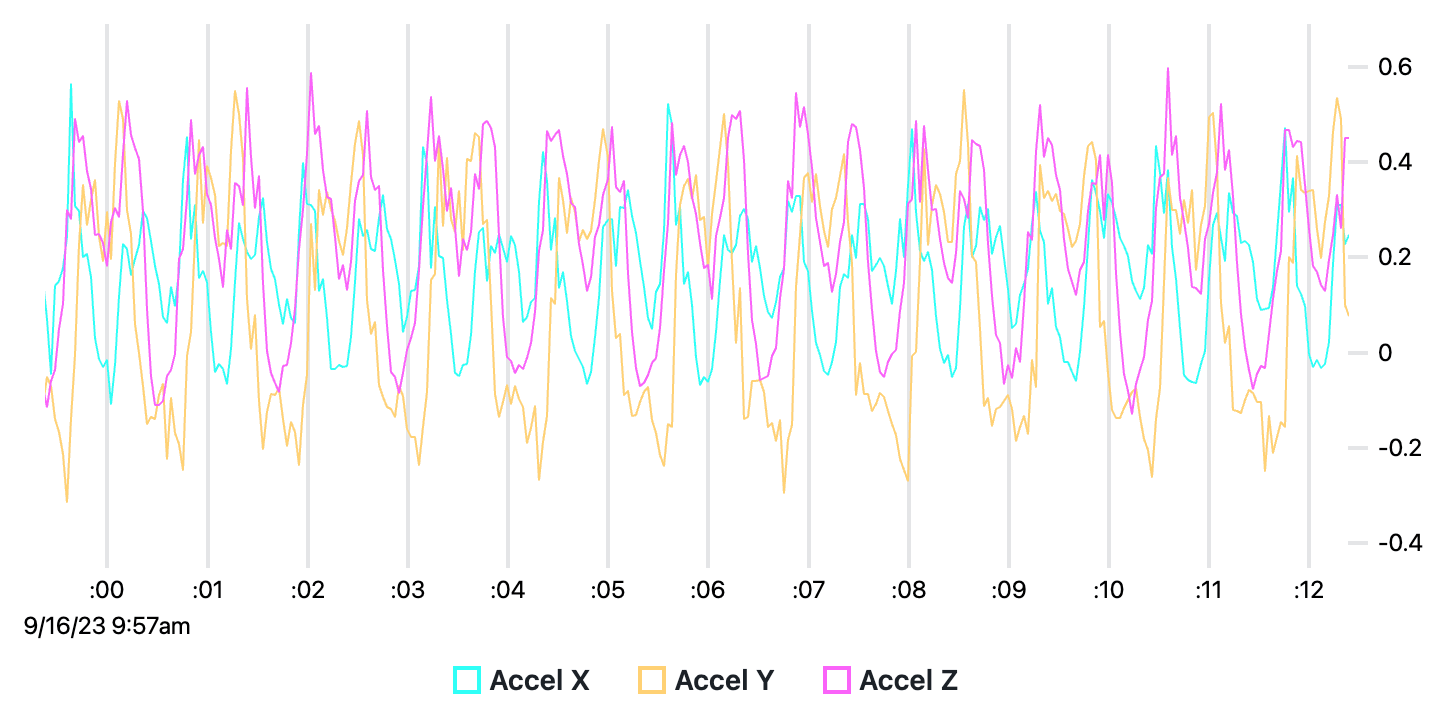}
            \caption{Acceleration data from an uphill section.}
            \label{fig:accel-uphill}
    \end{figure}

Biathlon and any other movements can be modeled as sections of movement that can be divided into smaller and smaller subsections. The macro-view (see Figure \ref{fig:racetrack-schema}) tells us the shooting sequence consists of entering the shooting range followed by a shooting period and ending by exiting the shooting range and starting running the uphill section. If the uphill section is divided into multiple subsections, the analysis can be done at a micro level (refer to Figure \ref{fig:accel-uphill}). It shows the precise sequence of motion to complete one uphill step. First, the arms of the athletes are moved forward, then the upper body leans forward, and the feet follow. This sequence always happens in the same order but with different forces and duration. Such insight gained represents the knowledge of this movement and should be encoded in a directed graph. 
    
After a brief analysis of motion in biathlon, the sequence of movements and their temporal dependency emerge as perfect properties for describing motion sequences. The simplest movement consists of two actions (i.e., referred to as points of interest) and can be noted as $A \rightarrow B$, which means $A$ is a condition for $B$ (or simply said, $A$ needs to happen before $B$). Taking the example from above, one can write $UE \rightarrow UL$ where $UE$ = uphill enter and $UL$ = uphill leave/exit, meaning that exiting the uphill section could only happen after the uphill section was entered in the first place. This way, all temporal dependencies are simply defined as:    

\begin{tabbing}
\hspace{2cm} \= \kill
   Definitions: \\
$S$  \> Start \\
$UE$ \> Enter uphill \\
$UL$ \> Exit/leave uphill \\
$P$  \> Penalty round \\
$F$  \> Finish \\
$RE$ \> Enter shooting range \\
$SS$ \> Start shooting \\
$SF$ \> Finish shooting \\
$RL$ \> Leave/exit shooting range \\
\end{tabbing}

Dependencies:
\begin{align*}
  & S \rightarrow UE && \text{(Start to entering uphill)} \\
  & UE \rightarrow UL && \text{(Entering uphill to exiting uphill)} \\
  & UL \rightarrow RE && \text{(Exiting uphill to entering shooting range)} \\
  & UL \rightarrow F  && \text{(Exiting uphill to finish line)} \\
  & RE \rightarrow RL && \text{(Entering shooting range to leaving shooting r.)} \\
  & RE \rightarrow SS && \text{(Entering shooting range to start shooting)} \\
  & SS \rightarrow SF && \text{(Start shooting to finish shooting)} \\
  & SF \rightarrow RL && \text{(Finish shooting to leaving shooting range)} \\
  & RL \rightarrow P  && \text{(Leaving shooting range to penalty round)} \\
  & P \rightarrow P   && \text{(Penalty round to another penalty round)} \\
  & P \rightarrow UE  && \text{(Penalty round to entering uphill)} \\
\end{align*}

    Those dependencies can be placed into one of the most common and well-studied data structures in computer science known as graphs \cite{West2000-ho}. In this case, directed graphs will be utilized to encode the temporal dependencies. A directed graph is defined as an ordered pair $G = (V, E)$ where
    
    \begin{itemize}
    \item $V$ is a set whose elements are called vertices or nodes. The representations will be used to depict actions or points of interest, such as reaching a specific speed or experiencing acceleration in a particular direction.
    \item $E$ is a set of ordered pairs of vertices called directed edges. The representations will be used to illustrate the temporal dependencies of the points of interest mentioned in the example with macro and micro-views in biathlon.
    \end{itemize}
    
    \begin{figure}[h]
        \centering
        \includegraphics[width=\columnwidth]{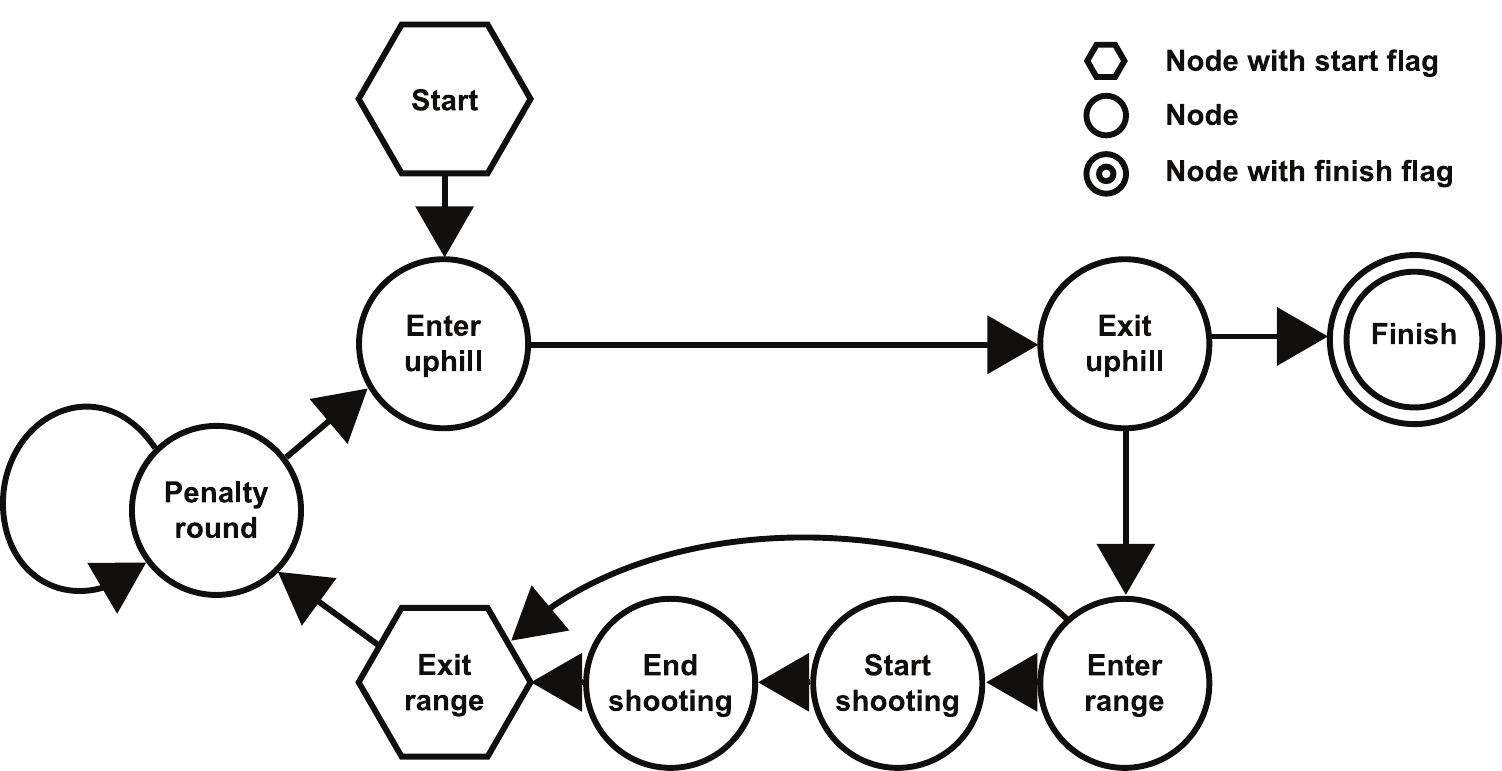}
        \caption{Macro-view of biathlon. Turning sequence of events into a directed graph.}
        \label{fig:macro-view-biathlon}
    \end{figure}
    
    Nodes represent specific points of interest, like entering or exiting an area or the start of a specific movement. The directed edges encode the order depending on time. In this manner, any sequence and, thus, any kind of human activity can be encoded into a data structure well understood by computers.
    
\subsection{Detection of points of interest in multi-sensor datasets} \label{detection}
    
    With datasets consisting of millions of data points, it is very inefficient to analyze each data point. The goal is to identify specific points in time that signal the occurrence of events required to detect a more complex motion. Given that most wearables record data from various sensors, it is important to develop methods for detecting both generic events and sensor-specific events. The most common sensor consists of an inertial measurement unit (IMU) sensor (which delivers acceleration and angular velocity data) and a global navigation satellite system (GNSS) receiver (which delivers position and speed data). The following triggers are helpful to detect points of interest in big data sets:
    
    \subsubsection{Generic triggers}
    
    \begin{itemize}
    \item Edge detection: Falling, rising, or change (both rising and falling) edge commonly used in signal processing. Used in devices like oscilloscopes to detect events.
    \begin{figure}[h]
        \centering
        \includegraphics[width=\columnwidth]{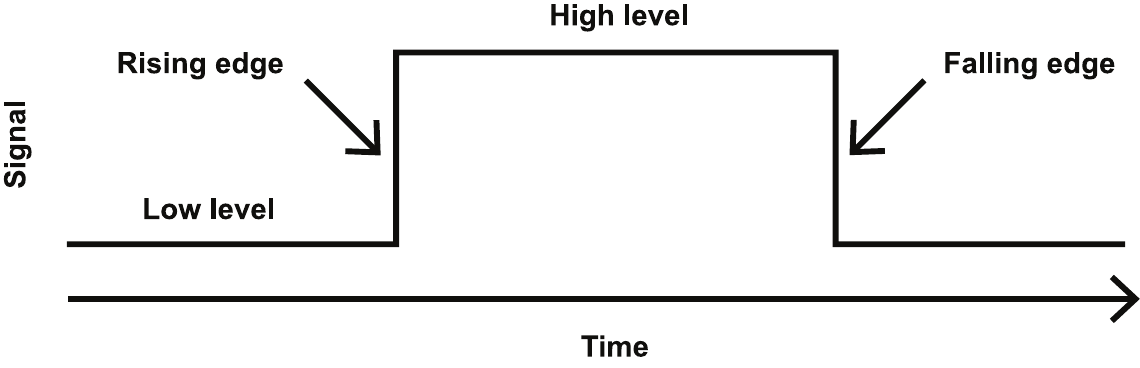}
        \caption{Edge detection anatomy.}
    \end{figure}
    The simplest implementation looks at a specific threshold. Mathematically, this function can be described as:
    \[ f(x) = \begin{cases} 
          1 & x\geq threshold \\
          0 & otherwise 
       \end{cases}
    \]
    \item Peak detection: Detects peaks by comparing local minima/maxima with neighboring values.
    \begin{figure}[h]
        \centering
        \includegraphics[width=\columnwidth]{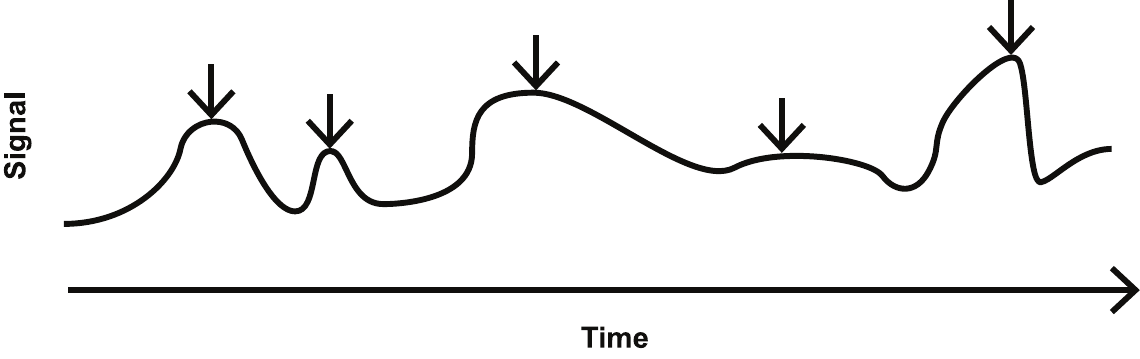}
        \caption{Peak detection anatomy.}
    \end{figure}
    There are many different algorithms to detect peaks. Some relevant examples are focused on electrocardiography \cite{Scholkmann2012}, and there are even specialized ones \cite{Koka2022} on detecting heartbeats. Nevertheless, we will not go into details about their inner workings as this is out of the scope of this work. Note that generic triggers can be used on any numerical dataset.
    \end{itemize}
    
    \subsubsection{Sensor-specific triggers}
    \begin{itemize}
    \item Location-based detection: Detects when a line (virtual gate) is crossed (intersection), or areas are entered/exited. Note that sensor-specific triggers can only be used on specific datasets.
    
    \begin{figure}[h]
        \centering
        \includegraphics[width=\columnwidth]{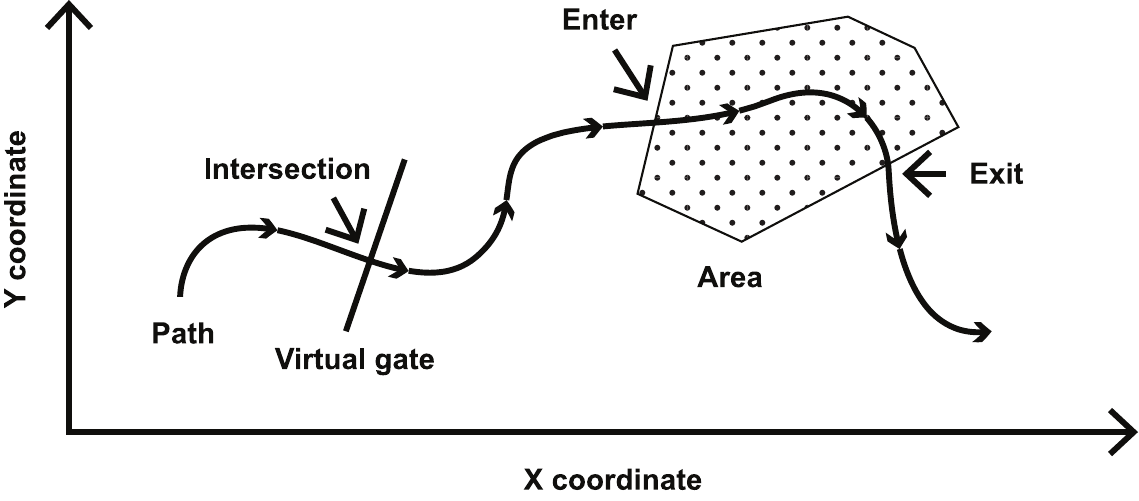}
        \caption{Location-based detection anatomy.}
    \end{figure}
    
    There are multiple ways to calculate intersections. It depends on the given coordinates. In a Cartesian coordinate system, the two line equations $y=ax+c$  and $y=bx+d$ are checked with simple rearranging and substitutions if and where those lines intersect. The complexity increases when considering the geographic coordinate system used in GNSS coordinates. Hence, it is essential to consider the model used to represent Earth. The Earth is commonly depicted as an ellipsoid, often with the WGS84 model, which is also used by the Global Positioning System (GPS) \cite{Fell2001}. For example, a simple formula to calculate the distance on a sphere is the Haversine formula \cite{Van_Brummelen2017}. One can also project the geographic coordinates to a Cartesian coordinate system and do calculations as mentioned above, but this only works for small distances.
    \end{itemize}
    
\begin{figure}
        \centering
    \includegraphics[width=\columnwidth]{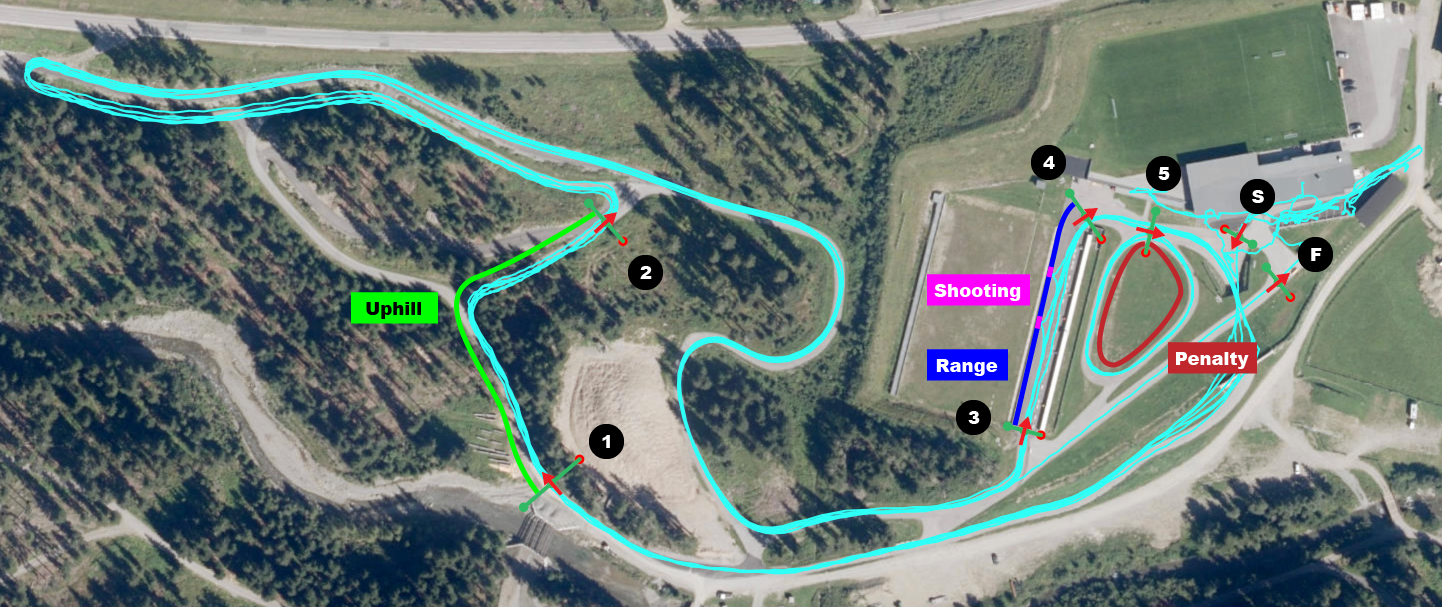}
    \caption{Virtual gates track layout.}
    \label{fig:track-layout-virtual-gates}
\end{figure}

   To measure a typical biathlon race, location-based detections, often referred to as virtual gates, are utilized. To better illustrate how the proposed algorithm works, the following sample data set will be used.  The virtual gate $S$ in  Figure \ref{fig:track-layout-virtual-gates} marks the start line. Next will be the uphill section consisting of gates $1$ and $2$. Then, athletes will enter the range enclosed by gates $3$ and $4$. The shooting section in the range will be detected based on the measurement of the speed when lying down (lower than 1m/s) and getting back up again (higher than 1m/s). Then, for each missed shot out of 5 total shots, athletes need to run the penalty round. So, if an athlete misses 2 out of 5, he must do two penalty rounds. Gate number $4$ is special in that it marks the beginning/ending of a lap. In total, six laps were done. Shooting is only done every 2\textsuperscript{nd} lap.\\

A point of interest (POI) refers to a specific moment in time (timestamp) of an event that is being measured. For example, gate 1 will produce a POI each time the athlete crosses the virtual line. The following notation is used for a POI: S\textsubscript{t} = point of interest $S$ (start) was triggered at timestamp $t$.

\begin{figure}
    \centering
    \includegraphics[width=\columnwidth]{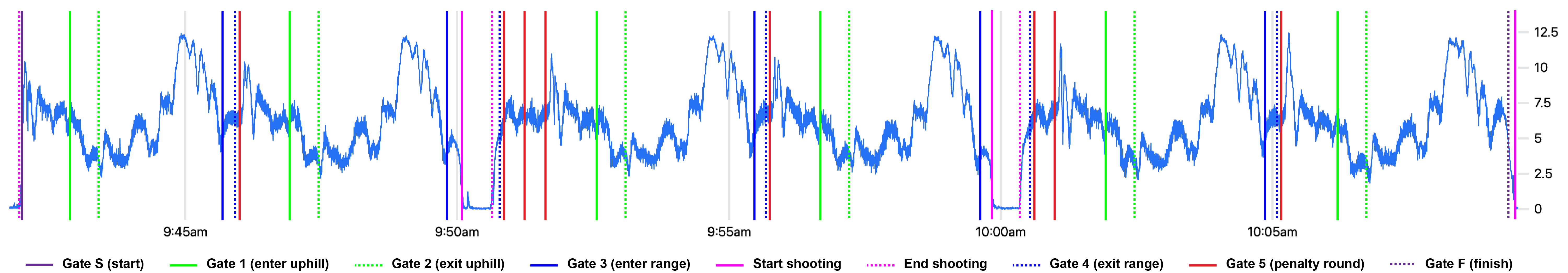}
    \caption{Triggers plotted in chart with speed data (m/s).}
    \label{fig:triggers-with-speed}
\end{figure}

Figure \ref{fig:triggers-with-speed} shows all POIs (vertical lines) plotted with athlete speed in m/s. All POIs are triggered by virtual lines except start/end shooting. The speed information triggers those. If athletes move more than 1m/s, it will trigger on the falling edge (start shooting) or rising edge (end shooting). A closer look at the first and last POI shows that those are "wrongly" start and end shooting triggers because triggers are unaware of any temporal dependency (as previously described).

\subsection{A direct graph to recognize complex movements}
\label{sec:recognize}

Up to now, the complex movement of interest is encoded into a directed graph and points of interest from our dataset. It is a matter of applying the graph to the points of interest to recognize our complex movement. In order to accomplish this, the graph needs to be traversed based on the points of interest. This problem can be translated into a Deterministic Finite Automaton (DFA) to explain this step more clearly. A DFA is described by a five-element tuple $(Q, \sum, \delta, q0, F)$ \cite{Hopcroft2000}:

\begin{tabbing}
\hspace{2cm} \= \kill 
$Q$  \>  states\\
$\sum$  \>  input alphabet\\
$\delta$  \>  transition functions\\
$q0$  \>  the starting state\\
$F$  \>  accepting states\\
\end{tabbing} 

The standard definition of $q0$ needs to be modified to allow for multiple starting states, considering the possibility of having multiple start nodes. A start node is defined as a node in our directed graph (as shown in Figure \ref{fig:macro-view-biathlon}) marked with a "start" flag. The following DFA is deduced from our example of biathlon:

\[DFA_{Biathlon} (Q, \sum, \delta, q0, F)\]

Firstly, $Q = \{S, UE, UL,  P, RE, RL, SS, SF, F\}$ represents all nodes of the directed graph. Secondly, $\sum = \{SF_{163}, S_{175}, UE_{214}, UL_{235}, …\} $ represent all points of interest from Section \ref{encoding}. 
Lastly, $\delta = \{S \rightarrow UE, UE \rightarrow UL, UL \rightarrow RE, UL \rightarrow F, RE \rightarrow RL, RE \rightarrow SS, SS \rightarrow SF, SF \rightarrow RL, RL \rightarrow P, P \rightarrow P, P \rightarrow UE\}$ represent all edges of the directed graph. To make it more readable,  the notation $A \rightarrow B$ is used to represent the transition function $\delta(A, Bt) = B$ where $A$ and $B$ represent a state and $At$ represents any POI triggered by $B$. All transition functions that are not mentioned will have no effect and can be defined as $\delta(Y, Xt) = Y$.  $q0 = \{S\}$ represent all nodes in the directed graph, which are start nodes. Furthermore,  $F = \{F\}$ represents all nodes in the directed graph that are finished nodes.

\begin{figure}
    \centering
    \includegraphics[width=\columnwidth]{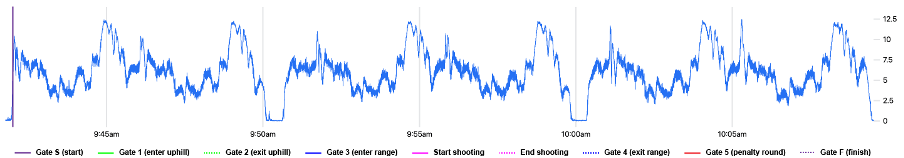}
    \caption{All points of interest from start nodes – in our example, only one.}
    \label{fig:poi-start}
\end{figure}

Starting points are defined as all points of interest triggered by a start node. Figure \ref{fig:poi-start} shows the POIs of the start node in the biathlon example. The presented Algorithm \ref{alg:start-points} describes this step.

\begin{algorithm}
\begin{algorithmic}
  \STATE $solutions \gets []$
  \STATE $startNodes \gets [node \mid node.start == \text{True}]$
  \FORALL{$s \in startNodes$}
  \STATE $poi \gets \text{fetchPointsOfInterestForNode}(s)$
  \FORALL{$p \in poi$}
  \STATE $sol \gets \text{findPartialSolution}(s, p)$
  \STATE $solutions.\text{append}(sol)$
  \ENDFOR
  \ENDFOR
  \RETURN $solutions$
  \caption{Searching for solutions starts at each start node.}
  \label{alg:start-points}
\end{algorithmic}
\end{algorithm}

The DFA runs until it hits an accepting state, leading to "a partial solution". A partial solution consists of the taken paths and timestamps and describes one detected movement. It is defined as only partial to ensure the detection of all movements in the dataset. The final or total solution will be built in the algorithm's last step and constructed from multiple partial solutions.
\begin{figure}
    \centering
    \includegraphics[width=\columnwidth]{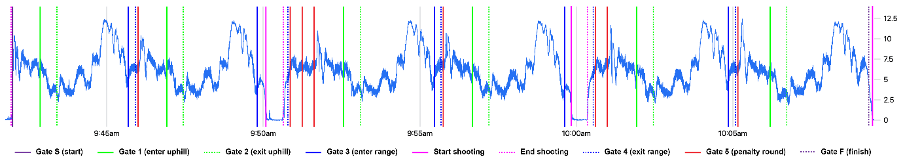}
    \caption{POIs from different nodes used as input alphabet for our DFA.}
    \label{fig:poi-all}
\end{figure}

After the automaton reaches an accepting state, the path taken and timestamps will be logged as a partial solution and then returned. The operation is described in Algorithm~\ref{alg:partial-solution}. Currently, each POI of a specific node is used as input and attempts to find a partial solution; however, there are specific cases where certain restrictions need to be applied. 

\begin{algorithm}
\begin{algorithmic}
  \STATE \textbf{Function} \textsc{findPartialSolution}($start, point, solution$)
  \STATE $solutions \gets []$
  \STATE $startNodes \gets [node \mid node.start == \text{True}]$
  \FORALL{$s \in startNodes$}
    \STATE $poi \gets \text{fetchPointsOfInterestForNode}(s)$
    \FORALL{$p \in poi$}
      \STATE $sol \gets \text{findPartialSolution}(s, p)$
      \STATE $solutions.\text{append}(sol)$
      \STATE \textsc{findPartialSolution}($start, point, solution$)
    \ENDFOR
  \ENDFOR
  \IF{$\text{len}(solutions) == 0$ \AND $\text{len}(solution) > 0$}
    \IF{$\text{lastNodeInSolution.finish} == \text{True}$}
      \RETURN $[solution]$
    \ENDIF
  \ENDIF
  \RETURN $solutions$
  \caption{The search for solutions starts at every starting node.}
  \label{alg:partial-solution}
\end{algorithmic}
\end{algorithm}

\begin{figure}
    \centering
    \includegraphics[width=\columnwidth]{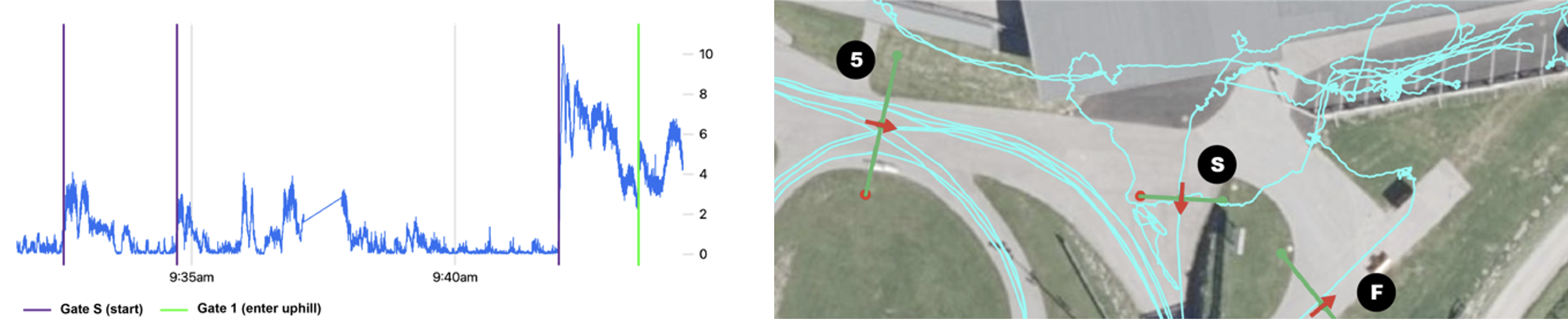}
    \caption{Start gate "S" was triggered multiple times. In this case, only the last start trigger is valid.}
    \label{fig:poi-start-detail}
\end{figure}

In our example, let's consider the gate labeled as "S" (start).
As shown in the map view of Figure \ref{fig:poi-start-detail}, the athlete crossed the virtual start gate $S$ several times before the race started. This might happen during the warm-up phase or the course inspection. Our approach would take those false triggers and build a valid solution. To avoid this scenario, users might specify a minimum or maximum duration for an edge. In this case, a maximum 60-second restriction is set for the path from gate $S$ (start) to gate 1 (enter uphill). It is a method to narrow down the solution space to only valid solutions according to training or race rules, reducing space and time complexity.

\subsection{Combine partial solutions to find multiple total solutions}
Only connected ones are found in the search for partial solutions, indicating a valid POI sequence followed by another sequence. Imagine a break between two sequences: One partial solution will be found for the first sequence and another for the second sequence.

\begin{figure}
    \centering
    \includegraphics[width=0.9\columnwidth]{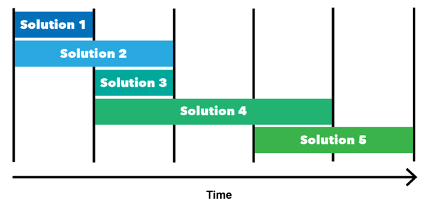}
    \caption{Five partial solutions displayed by their temporal coverage. Note that each color represents a different partial solution.}
    \label{fig:partial-solutions}
\end{figure}

All partial solutions should be combined in all possible valid ways. "Valid" means that overlapping sequences cannot happen simultaneously. Temporal order must be preserved, and solutions cannot be moved on the time axis. A list of solutions will be obtained, while a single solution may consist of one or multiple partial solutions. Each partial solution is also considered a valid solution.

\begin{figure}
    \centering
    \includegraphics[width=0.9\columnwidth]{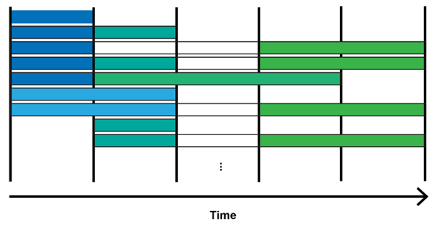}
    \caption{Combining partial solutions to form valid total solutions. }
    \label{fig:partial-solution-combinations}
\end{figure}

\subsection{Find optimal total solution}
After identifying all potential solutions, establishing a metric for comparing and ranking them is essential. The specific metric will vary depending on the particular use case, but reasonable assumptions can be made to strike a balance between different approaches. Let's examine some of these approaches:

\subsubsection{Maximize number of partial solutions}
The most obvious would be maximizing the number of partial solutions in one solution. Let $p_i$ be the partial solution of $s$ with index $i$ and $n$, which is the partial solution in $s$. The solution $s$ is ranked by:

\[rankBy(s) = \sum_{i=0}^n 1\]

\begin{figure}[h]
    \centering
    \includegraphics[width=0.9\columnwidth]{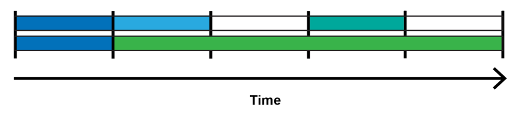}
    \caption{Fragmented solution vs. de-fragmented solution.}
    \label{fig:solution-frag-defrag}
\end{figure}

The disadvantage would be that many short solutions would be ranked at the top. This leads to a fragmented solution shown in Figure \ref{fig:solution-frag-defrag}.

\begin{figure}
    \centering
    \includegraphics[width=0.9\columnwidth]{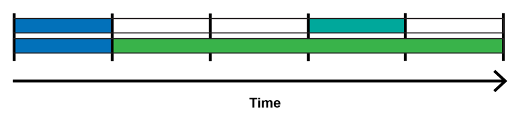}
    \caption{Two solutions with each two partial solutions but different total duration.}
    \label{fig:solution-rank-non-deter}
\end{figure}

Fragmented solutions tend to lead to wrong solutions as false triggers in Section \ref{detection} are promoted. This is because maximizing the number of partial solutions minimizes the duration covered by each partial solution. The algorithm in Section \ref{sec:recognize} produces partial solutions with short coverage (i.e., especially when false triggers occur).
In general, there is no way of determining the "correct" trigger. It might be the first one, but it could also be the second one. Suppose multiple solutions have the same number of partial solutions. In that case, the result will be an undefined behavior as it's impossible to determine the "best" solution (as shown in Figure \ref{fig:solution-rank-non-deter}).

\subsubsection{Maximize the covered duration}
To mitigate a fragmented solution, the total duration is maximized and covered in the solution. The total duration of a solution is the sum of all durations of partial solutions. Let $duration$ be a function to calculate the duration (end time – start time), and then the total duration is the sum of it. The solution $s$ is ranked by:

\[rankBy(s) = \sum_{i=0}^n duration(p_i)\]

\subsubsection{Combination}
\begin{figure}
    \centering
    \includegraphics[width=0.9\columnwidth]{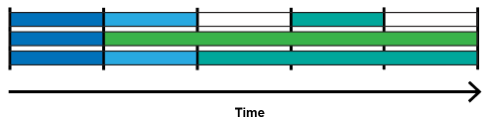}
    \caption{Last solution will be taken.}
    \label{fig:solution-rank-combination}
\end{figure}

Both approaches will inevitably lead to non-deterministic behavior as there will be a lot of solutions with the same number of partial solutions or covered duration. The best approach is to combine both methods. First, rank for the maximum covered duration and then maximize the number of partial solutions(as illustrated in Figure \ref{fig:solution-rank-combination}).

\section{Evaluation}
\label{chap:experiments}

\subsection{Implementation, Testbed, and Dataset} 
\label{chap:prelim}
In this paper, sample data was collected in biathlon by a wearable IMU sensor and GNSS receiver. The sensor "OCULUS" is made by the Austrian company Lympik\footnote{Lympik, (\url{https://www.lympik.com})}, a sports technology company focused on professional sports analytics. The accelerometer was configured at 50Hz and capable of measuring up to 16G (i.e., gravitational force). The gyroscope was also configured at 50Hz and an upper limit of 2000 degrees per second. The GNSS receiver was configured to record at 10Hz in multi-constellation (i.e., GPS, Galileo, GLONASS) mode, and Satellite-based Augmentation Systems (SBAS) were also enabled.

\begin{figure}[h]
    \centering
    \includegraphics[width=1\linewidth]{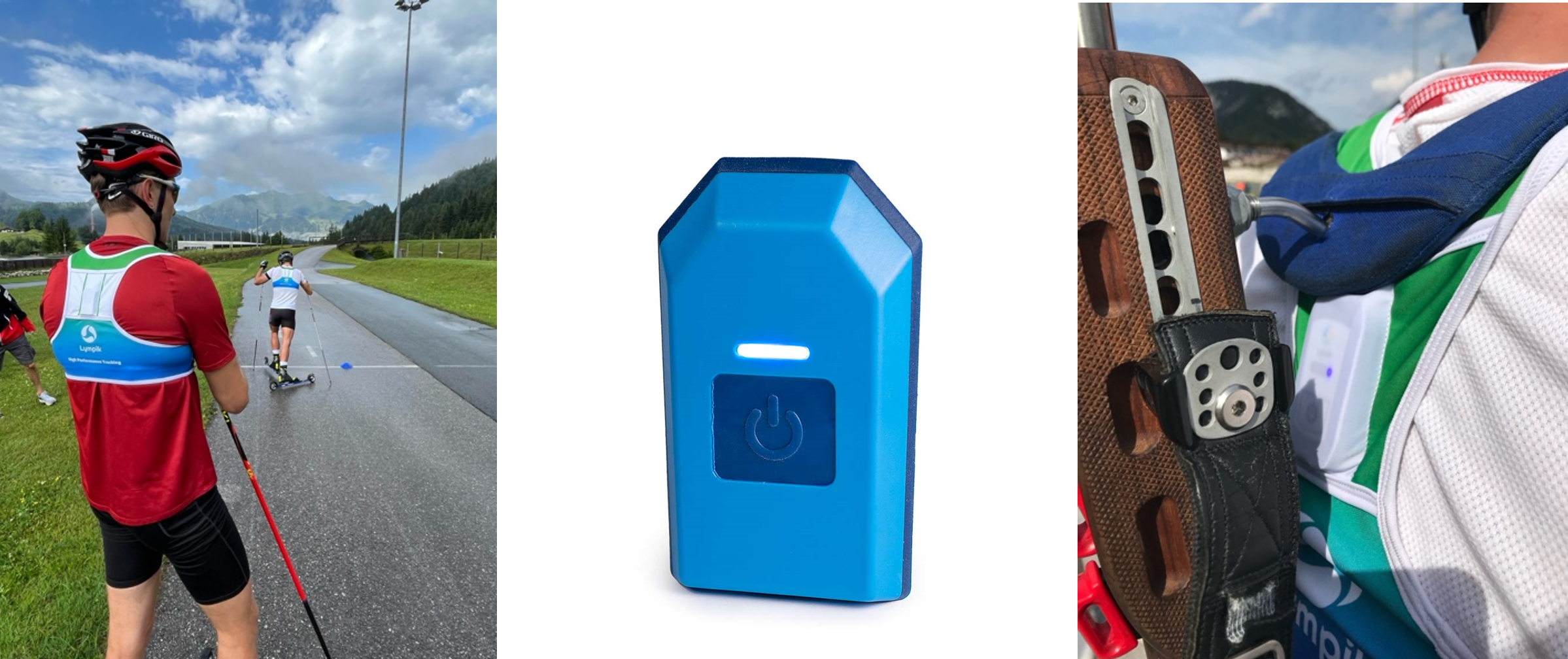}
    \caption{OCULUS tracking device mounted on biathlon athlete.}
    \label{fig:oculus}
\end{figure}

Our test subjects were three professional athletes. The data was recorded in the summer when the biathlon was simulated on special rollers. The wearable was mounted on the athlete with a special shirt while the rifle was carried above the sensor to avoid disturbing it while taking it off and picking it up. 

\begin{figure}[h]
    \centering
    \includegraphics[width=1\linewidth]{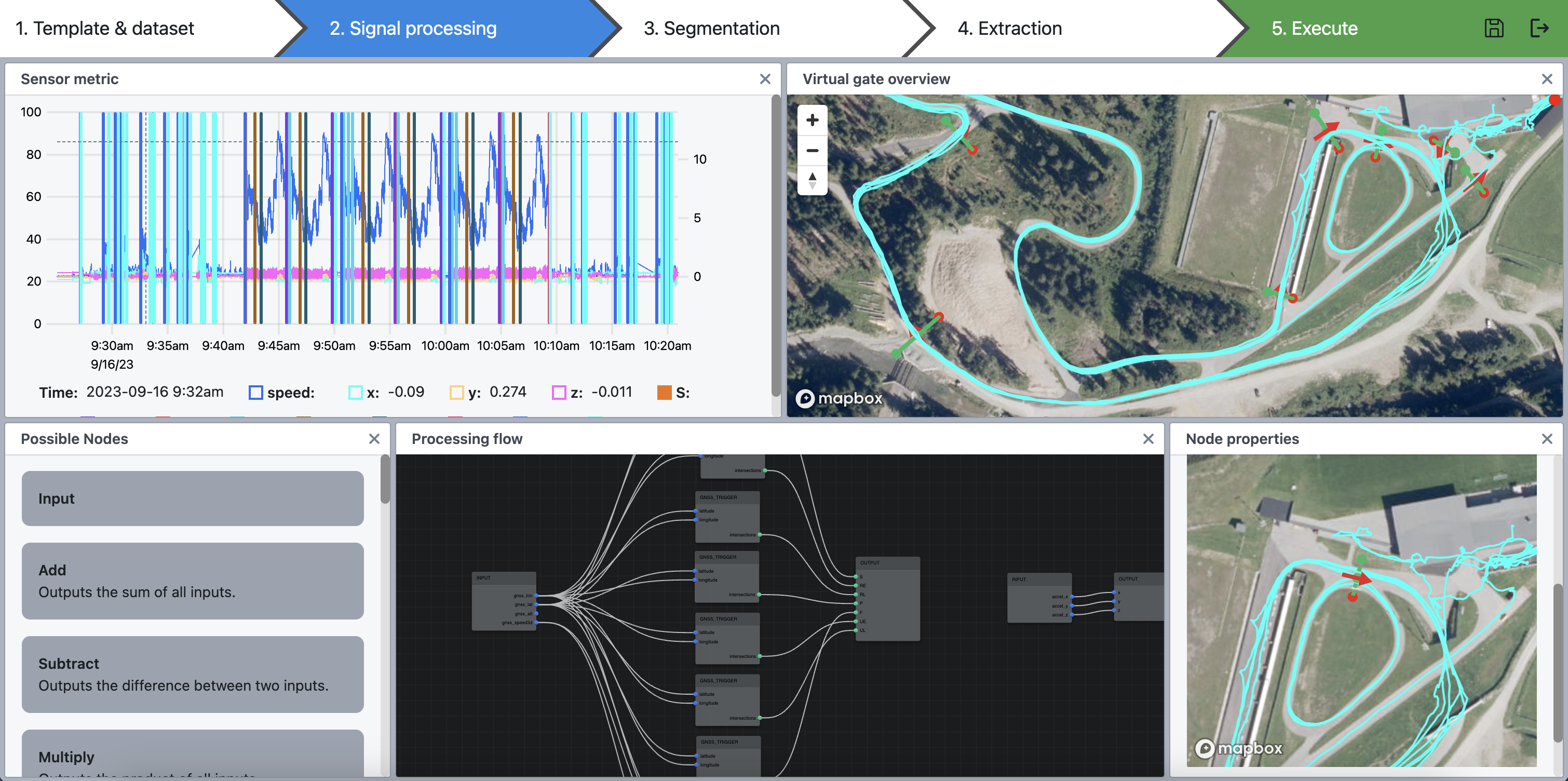}
    \caption{Lympik sensor studio for analyzing sensor data.}
    \label{fig:sensor-studio}
\end{figure}

Data analytics and visualization were done in Lympik Sensor Studio$^1$, a software service that quickly analyzes multi-sensor data. All functions used in the Studio are easily reproducible with simple scripts. {The test implementation was done using Python. The raw binary sensor data was converted to standard units and signal processing was done with Scipy \footnote{\url{https://scipy.org/}}. The location-based trigger is based on a simple line intersection in an Euclidean system.}

For the detection of POI in our datasets, two kinds of triggers are used: Edge detection based on speed (i.e., the threshold was set to 1m/s) for detecting shooting start/finish and location-based triggers, which were positioned as shown in Figure \ref{fig:track-layout-virtual-gates}. The domain-specific knowledge was encoded into a graph as presented in Figure \ref{fig:macro-view-biathlon}. The resulting POIs were applied to our DFA as described in Section \ref{sec:recognize}. The path taken was recorded, and each edge was color-coded to recognize each segment easily. The segmentation is shown in a bar chart where the x-axis represents the time axis.

\subsection{Experiments and Results}
 
Our experiments were done with three biathlon athletes.  This experiment aimed to test our approach to handling large datasets. Each dataset contains around 200k GNSS data points and 500k IMU data points\footnote{\url{https://doi.org/10.5281/zenodo.13208678}}. The IMU of one tracker was configured at 200hz to test different dataset sizes as well. The knowledge of the track was obtained from a professional trainer and encoded into the graph as described in Section \ref{encoding}.

In total, each athlete did six laps. Shooting was done only every 2\textsuperscript{nd} lap; so, the athletes passed the shooting range without shooting in the 1\textsuperscript{st}, 3\textsuperscript{rd} and 5\textsuperscript{th}. In the 6\textsuperscript{th} and final lap, the athletes also skipped the shooting range and went to the finish. The raw GNSS data and virtual gates are visualized on a map. The chart displays the POIs triggered by each virtual gate, and the bar chart presents the optimal solution generated by the algorithm.

\begin{figure}
    \centering
    \includegraphics[width=1\linewidth]{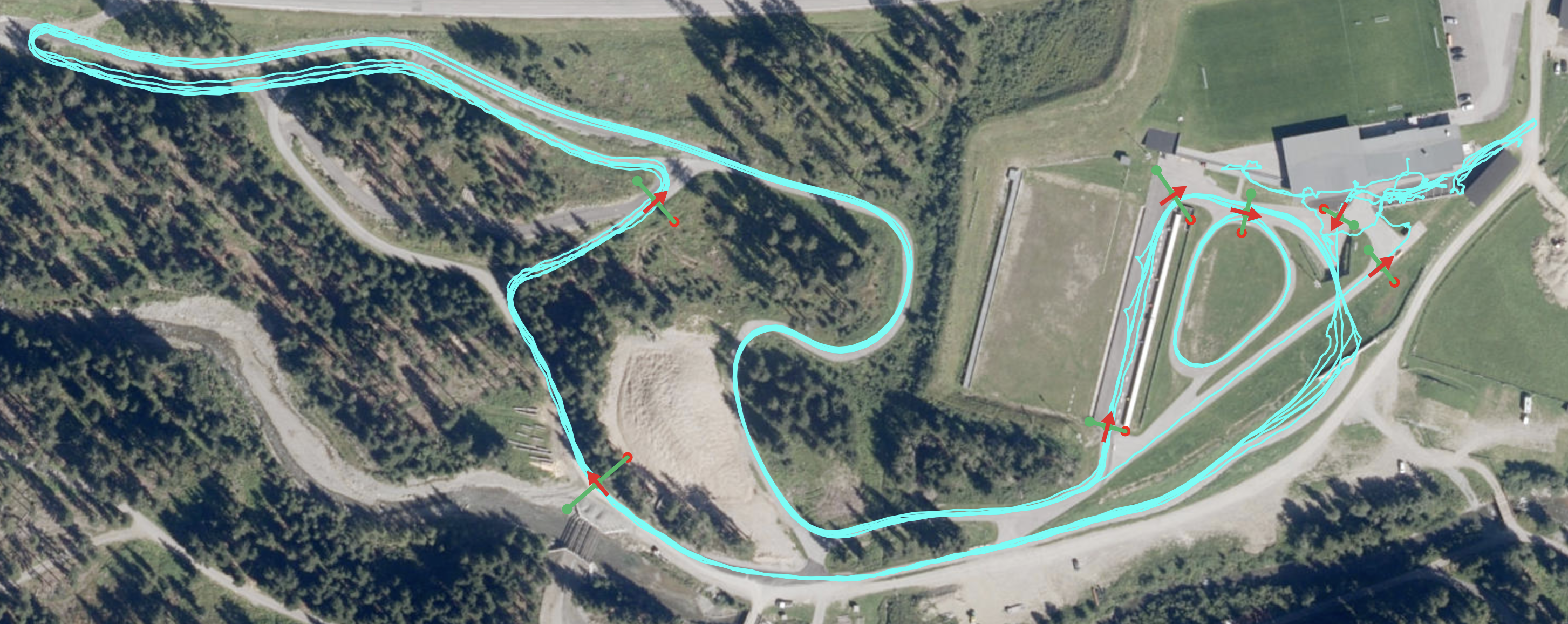}
    \includegraphics[width=1\linewidth]{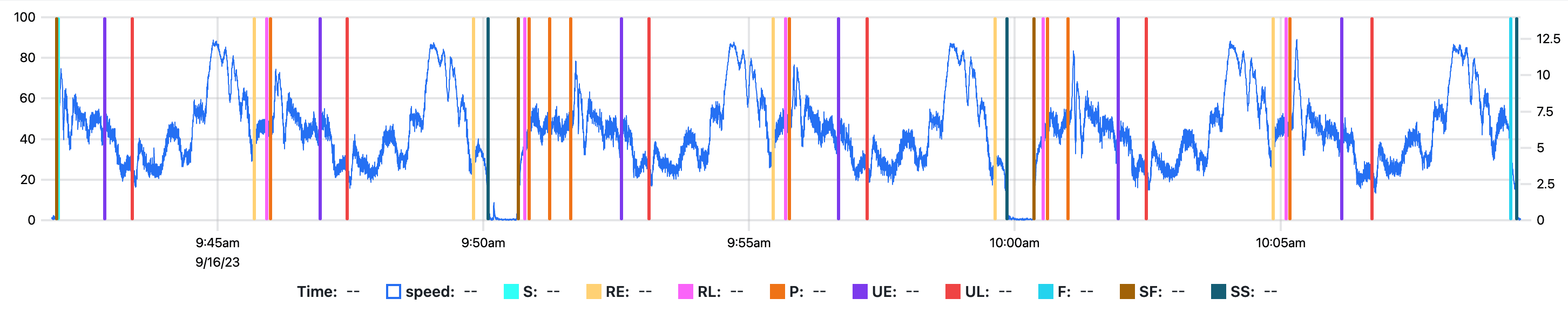}
    \includegraphics[width=1\linewidth]{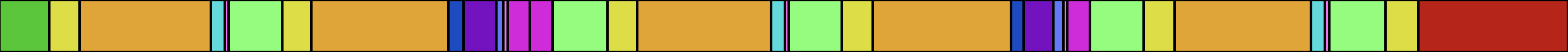}
    \includegraphics[width=1\linewidth]{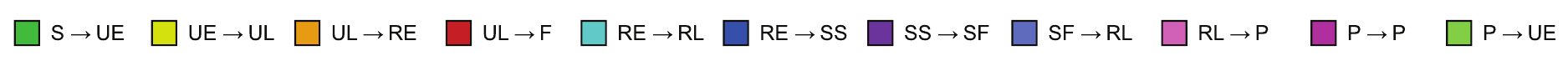}
    \caption{Athlete A - dataset 191.9k GNSS data points, 503.1k IMU data points.}
    \label{fig:exp-a}
\end{figure}

\begin{figure}
    \centering
    \includegraphics[width=1\linewidth]{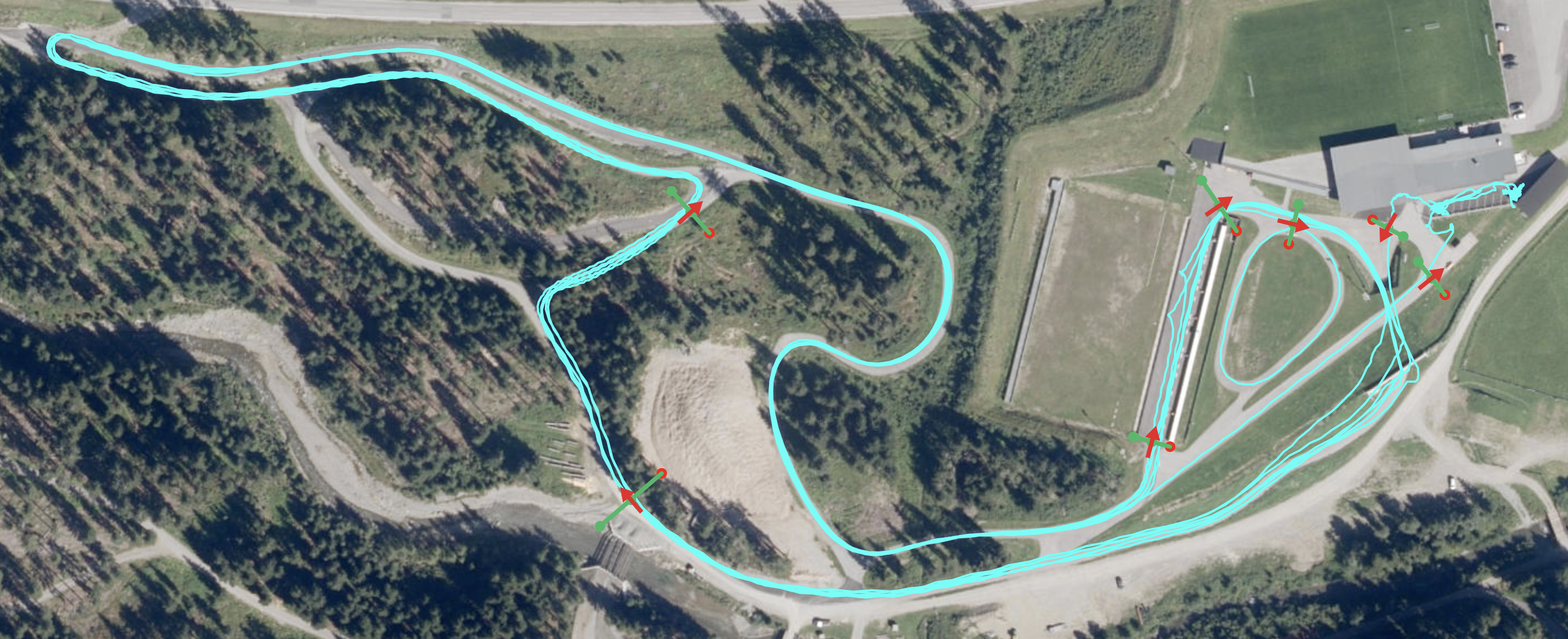}
    \includegraphics[width=1\linewidth]{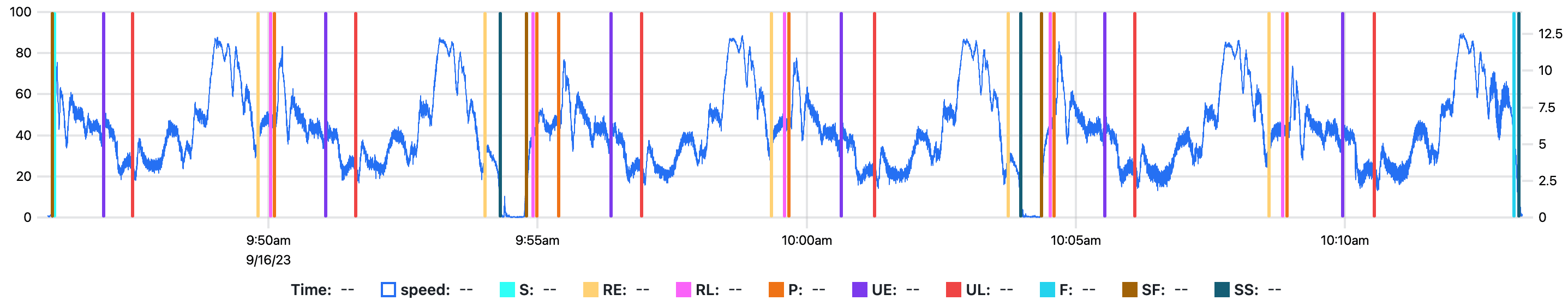}
    \includegraphics[width=1\linewidth]{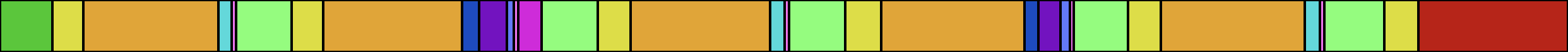}
    \includegraphics[width=1\linewidth]{graphics/poi-legend.pdf}
    \caption{Athlete B - dataset 197.6k GNSS data points, 493.7k IMU data points.}
    \label{fig:exp-b}
\end{figure}

\begin{figure}
    \centering
    \includegraphics[width=1\linewidth]{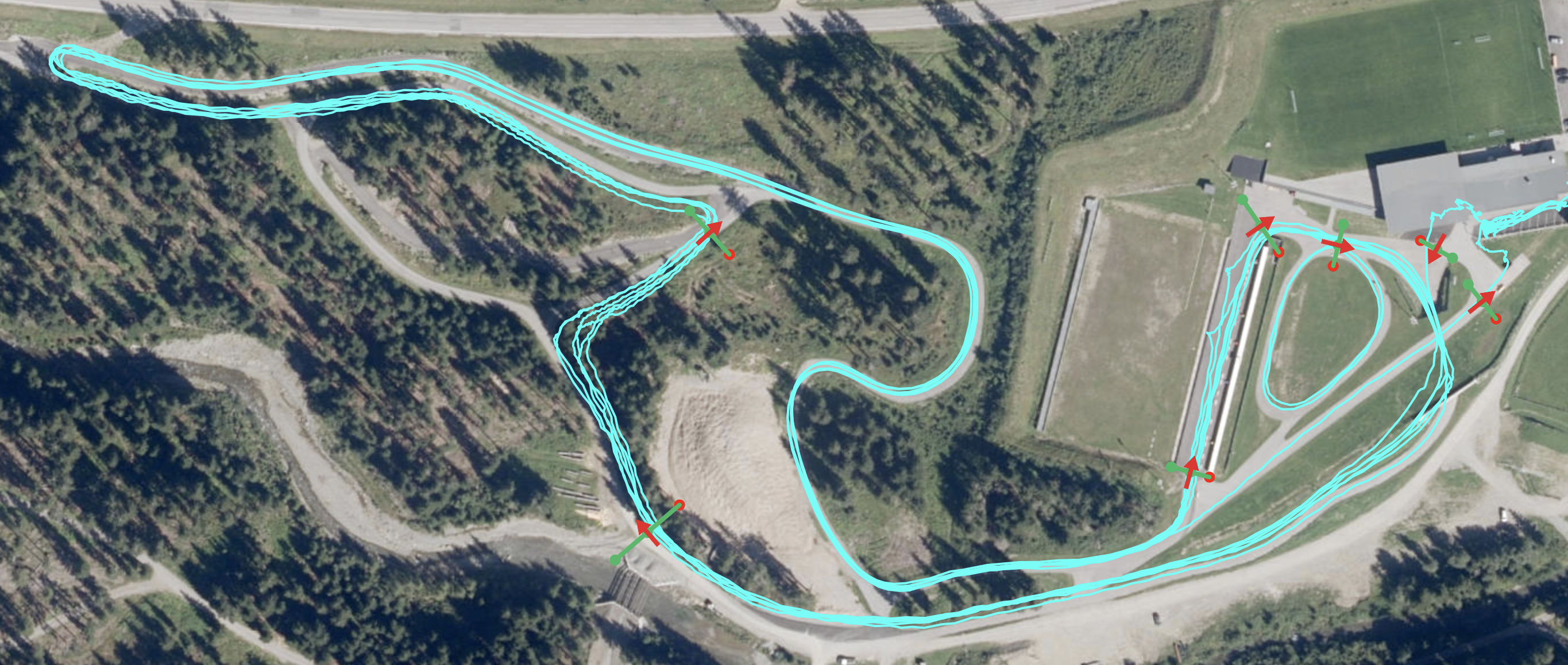}
    \includegraphics[width=1\linewidth]{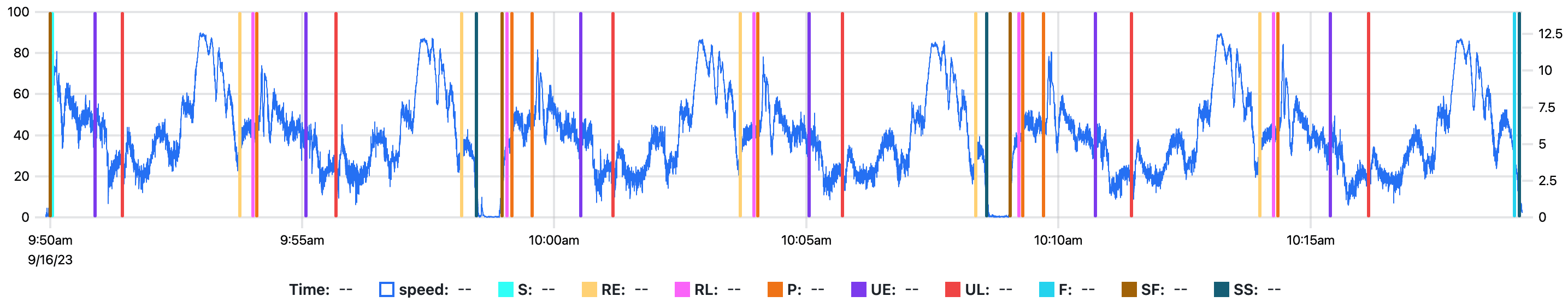}
    \includegraphics[width=1\linewidth]{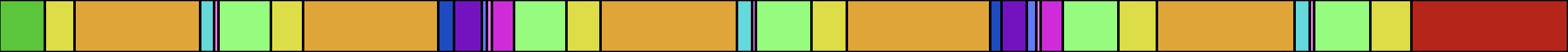}
    \includegraphics[width=1\linewidth]{graphics/poi-legend.pdf}
    \caption{Athlete C - dataset 205.5k GNSS data points, 2.1M IMU data points.}
    \label{fig:exp-c}
\end{figure}

{All three datasets were correctly segmented and found the optimal solution. Based on each segment, further calculations were made, such as calculating the average speed or maximum forces applied to the athlete. In dataset A (see Figure \ref{fig:exp-a}), the bar chart clearly shows the number of penalty rounds the athlete takes in pink $P \rightarrow P$. In the first shooting, the athlete missed two shots and took two penalty rounds, and in the 2\textsuperscript{nd} shooting missed one. Athlete B, shown in Figure \ref{fig:exp-b}, did one penalty round in the first shooting and none in the 2\textsuperscript{nd}. Results from Athlete C in Figure \ref{fig:exp-c} show that the athlete missed one shot each time.}

{Upon closer examination, the numerical data in Table \ref{tab:exp-shooting} reveals an interesting observation. All athletes show a faster range time ($RE \rightarrow SS \rightarrow SF \rightarrow RL$) in the 4\textsuperscript{th} lap than the 2\textsuperscript{nd}. Also, the Z-acceleration while shooting is different. The easy explanation for this correlation is that in the 2\textsuperscript{nd} lap, athletes were told to shoot in a lying position while in the 4\textsuperscript{th}, they had to stand. Standing results in a faster range time overall and different acceleration information.}

\begin{table}
  \centering
  \begin{tabular}{|c||c|c|c|c|}
    \hline
    Dataset & Lap & Range time & Shooting time & Shooting Z-accel.\\
    \hline
    \hline
     Athlete A & 2 & 57.53s & 33.90s & 0.33G \\
     Athlete A & 4 & 54.49s & 30.80s & -0.03G \\
    \hline
     Athlete B & 2 & 53.47s & 29.00s & 0.27G \\
     Athlete B & 4 & 46.85s & 22.90s & -0.07G \\
    \hline
     Athlete C & 2 & 53.67s & 30.60s & 0.49G \\
     Athlete C & 4 & 51.11s & 28.10s & 0.18G \\
    \hline
  \end{tabular}
  \caption{Duration and acceleration measured in shooting range.}
  \label{tab:exp-shooting}
\end{table}

\begin{figure}
    \centering
    \includegraphics[width=1\columnwidth]{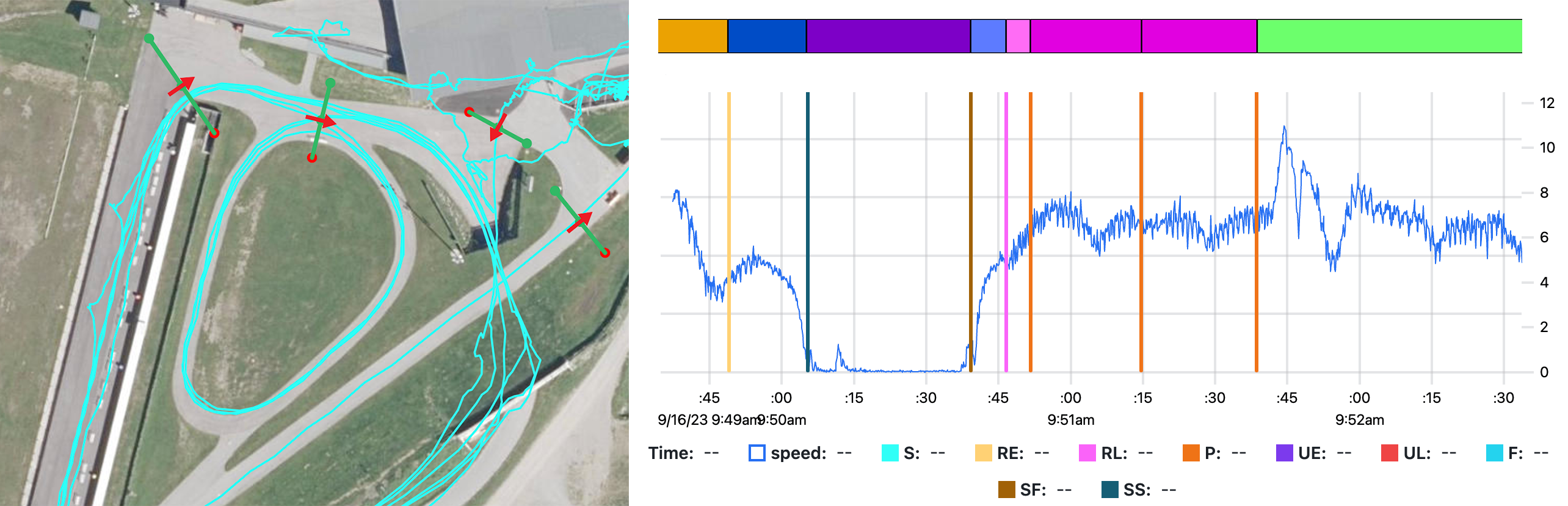}
    \caption{A detailed view of the penalty round.}
    \label{fig:exp-penalty}
\end{figure}

{Another experiment was to test the penalty round, which marks a special case. In this case, the athlete crosses multiple times (i.e., depending on the missed shots) the same line, and the segmentation still needs to work correctly (i.e., the penalty gate trigger in orange in the line chart). As shown in Figure \ref{fig:exp-penalty}, the three vertical lines indicate that the athlete crossed this gate three times. One can see the penalty was also correctly segmented (i.e., the bar chart in pink). The bar chart segmentation aligns perfectly with the line chart representing the POIs in this example taken from athlete A's dataset.}

\subsection{Discussion, Limitations, and Future Work}

All results presented in Section \ref{chap:experiments} were calculated on datasets with a few hundred thousand data points within 1-2 milliseconds on a one vCPU machine on the Google Cloud Platform with 2GiB of memory. Compared to other solutions utilizing neural networks (i.e., examples mentioned in the introduction), this method uses well-optimized and highly efficient analytic approaches combined with domain-specific knowledge to detect complex movements in a fast and resource-efficient way. {There was no significant difference in processing time between datasets (i.e., dataset C had four times more IMU data than the others), further solidifying our approach for big datasets.} Furthermore, no big datasets for training are needed, and traceability is a further advantage of the proposed approach. 

The limitation of the proposed approach is the required domain-specific knowledge. One needs to know how a complex movement can be split into points of interest. A further challenge is posed by movement sequences, in which athletes do not perform the same way every time. In the future, we plan to parallelize the search for sub-solutions in Section \ref{sec:recognize}. One potential area to explore is the potential to remove the necessity for domain-specific knowledge. This could be achieved by utilizing a neural network or federated learning concepts (i.e., as presented in \cite{murturi2023community}) to embody domain-specific insights from extensive datasets and construct a graph based on the acquired knowledge. Future work remains investigating possibilities for real-time insights while enabling data processing in a distributed manner in the computing continuum \cite{donta2023exploring,info14030198}. Lastly, we will explore integrating advanced sensor data analytics with AI planning techniques \cite{murturi2022utilizing} in edge computing environments \cite{murturi2022decent} to enhance the real-time performance and scalability of physical activity monitoring systems.    

\section{Conclusion}  
\label{conclusion}
 Integrating advanced wearable sensor devices has revolutionized capturing detailed movement data during physical activities. Such capability provides invaluable insights into performance metrics, learning fatigue points, and tracking movement efficiency. The proposed approach leverages sensor data and, via graph-based, captures and analyzes detailed movement data from individuals during various physical activities. This methodology offers a comprehensive solution for identifying critical performance metrics and fatigue points. Visualizing these data through graphs enables a clear and intuitive understanding of where and why performance drops occur.  Nevertheless, this paper outlines just the initial stage of operationalizing the framework. In future work, we aim to develop a comprehensive technical framework and provide a thorough evaluation.

\section*{{Acknowledgment}}
This work has been partially supported by the European Union's Horizon Europe research and innovation program under grant agreements No. 101135576 (INTEND) and No. 101070186 (TEADAL).

\bibliographystyle{ieeetr}
\bibliography{sample-base} 
\balance
\end{document}